\documentclass{aa} 
\usepackage{graphicx}
\usepackage{txfonts}
\begin{document}
   \title{Observational templates of star cluster disruption}

   \subtitle{The stellar group NGC~1901 in front of the Large
   Magellanic Cloud}

   \author {G. Carraro$^{1,2}$,
           R. de la Fuente Marcos$^{3}$, S. Villanova$^{1}$, C. Moni Bidin$^{2}$, C. de la Fuente Marcos$^{3}$, 
           H. Baumgardt $^{4}$, and  G. Solivella$^{5}$}

   \offprints{Giovanni Carraro: giovanni.carraro@unipd.it}

   \institute{$^1$Dipartimento di Astronomia, Universit\`a di Padova,
                  Vicolo Osservatorio 2, I-35122 Padova, Italy\\
              $^2$Departamento de Astronom\'ia, Universidad de Chile, 
	          Casilla 36-D, Santiago, Chile\\
              $^3$Suffolk University Madrid Campus, C/ Vi\~na 3, 
                  E-28003 Madrid, Spain\\
              $^4$AIfA, University of Bonn, Auf dem H\"ugel 71, 
                  D-53121 Bonn, Germany\\            
              $^5$Facultad de Ciencias Astron\'omicas y Geof\'{\i}sicas 
                  de la UNLP, IALP-CONICET, Paseo del Bosque s/n, La Plata, 
                  Argentina\\}

   \date{Received November 1, 2006; accepted XXXXXXX XX, XXXX}

 
   \abstract
       {Observations indicate that present-day star formation in the 
        Milky Way disk takes place in stellar ensembles or clusters 
        rather than in isolation. Bound, long lived stellar groups 
        are known as open clusters. They gradually lose stars and in 
        their final evolutionary stages they are severely disrupted 
        leaving an open cluster remnant made of a few stars.} 
       {In this paper, we study in detail the stellar content and 
        kinematics of the poorly populated star cluster NGC~1901. 
        This object appears projected against the Large Magellanic 
        Cloud. The aim of the present work is to derive the current 
        evolutionary status, binary fraction, age and mass of this 
        stellar group. These are fundamental quantities to compare 
        with those from $N$-body models in order to study the most 
        general topic of star cluster evolution and dissolution.}
       {The analysis is performed using wide-field photometry in 
        the UBVI pass-band, proper motions from the UCAC.2 
        catalog, and 3 epochs of high resolution spectroscopy,
        as well as results from extensive $N$-body calculations.}
       {The star group NGC~1901 is found to be an ensemble of solar 
        metallicity stars, 400$\pm$100 Myr old, with a core radius of 
        0.23 pc, a tidal radius of 1.0 pc, and located at 400$\pm$50 pc 
        from the Sun. Out of 13 confirmed members, only 5 single stars 
        have been found. Its estimated present-day binary fraction 
        is at least 62\%. The calculated heliocentric space motion of 
        the cluster is not compatible with possible membership in the
        Hyades stream.} 
       {Our results show that NGC~1901 is a clear prototype of open 
        cluster remnant characterized by a large value of the binary
        fraction and a significant depletion of low-mass stars. In the
        light of numerical simulations, this is compatible with NGC~1901
        being what remains of a larger system initially made of 
        500-750 stars.}


    \authorrunning{Carraro et al.}
    \titlerunning{The stellar group NGC~1901}

   \maketitle
%

\section{Introduction}
    Observations strongly suggest that present-day star formation in the 
    Milky Way disk takes place in stellar clumps rather than in isolation. 
    These stellar aggregates form out of giant molecular clouds and, in 
    principle, they can be born bound or unbound. Unbound, short lived 
    ($<$ 50 Myr) stellar ensembles are called associations; bound, long 
    lived stellar groups are known as open clusters. On the other hand, 
    simulations show that open clusters can also be formed out of the 
    remains of rich stellar associations (Kroupa et al. 2001). 

    As most of the field stars have been formed in the so-called clustered 
    mode (i.e., in clusters or associations), not in the dispersed mode 
    (i.e., in isolation), it is natural to identify these stellar clumps 
    as the {\it de facto} units of star formation in the Galactic disk 
    (Clarke et al. 2000). The idea of open clusters being fundamental units 
    of star formation is, however, controversial (Meyer et al. 2000) as it 
    has been argued that bound open clusters cannot contribute 
    significantly to the field star population because they are rare and 
    long lived (e.g. Roberts 1957). In contrast, most young embedded 
    clusters are thought to evolve into unbound stellar associations, which 
    produce the majority of stars that populate the Galactic disk (Lada \& 
    Lada 1991).

    $N$-body simulations show that the dynamical evolution of open star
    clusters is mainly the result of encounters among cluster members, 
    stellar evolution, encounters with giant molecular clouds, and the 
    effect of the Galactic tidal field. They gradually lose stars as they 
    evolve and in their final evolutionary stages they are disrupted, 
    leaving a cluster remnant (OCR, e.g. de la Fuente Marcos 1998) made of 
    a few stars. For a given Galactocentric distance, the distinctive 
    features of these ghostly objects depend upon the original membership 
    of the cluster, the fraction of primordial binaries (PBs), and the 
    initial mass function. The outcome of these numerical simulations shows
    a relatively large (about 20\%) percentage of binaries in OCRs of models 
    without PBs and up to 80\% for models with significant fraction of PBs 
    (30\%). The effects of preferential evaporation of low-mass stars in
    simulated OCRs are also quite important, with a significant depletion of 
    the primordial low-mass stellar (or substellar) population compared to 
    the group of more massive stars. The different evolutionary status 
    characteristic of classical open clusters and their remnants can be 
    traced in both their stellar composition and structural parameters, 
    particularly the degree of core-halo differentiation, that is larger in 
    evolved objects. Simulations can also provide information on the
    correlation between cluster life-times and their primordial properties,
    like membership. Averaging cluster life-times for entire sets of 
    numerical models (de la Fuente Marcos \& de la Fuente Marcos 2004)
    as a function of $N$ the following power law is obtained: 
    $\tau = 0.011 \ N^{0.68}$ (in Gyr) with a correlation coefficient of 
    $r = 0.995$ and associated errors in $\tau$ of $\sim$ 10\%. This slope 
    is very similar to the slopes found by Baumgardt (2001) and Baumgardt 
    \& Makino (2003) although the details of the simulations are rather 
    different. 

    Although the dynamical evolution of open star clusters has been well
    studied and our degree of understanding on the various evolutionary
    stages turning cluster members into field population is satisfactory,  
    the topic of detection of open cluster remnants is still controversial, 
    not only because their intrinsic observational properties make them 
    difficult to identify but also because there are no reliable 
    classification criteria. In fact, the main question still is, in general, 
    when does a star cluster become a cluster remnant?  A first attempt to 
    provide an answer to this important question can be found in Bica et al. 
    (2001). In their work, they assume that a cluster becomes significantly 
    depopulated and shows signs of being a remnant after losing 2/3 of its 
    initial population and they suggest the acronym POCR (Possible Open 
    Cluster Remnant) for a number of candidates. The application of this 
    criterion to simulated clusters is quite straightforward, but it is 
    difficult to use with real clusters mainly because star counts are 
    incomplete and there is no observational method to provide a reliable 
    estimate of the initial population of a given cluster. Unfortunately, 
    simulations and observational results both suggest that membership is 
    not the best criteria to determine if a dynamically evolved open cluster 
    is an OCR. In order to detect an OCR, certain contrast of the candidate 
    object against the stellar background is required. To be 
    identified initially as stellar overdensity, a cluster must have a 
    stellar density (or, more properly, a surface luminosity) larger than 
    the local Galactic value. Simulations suggest that, for poorly populated 
    open clusters, and before losing 2/3 of their initial population, the 
    stellar density has become lower than the surrounding field and the 
    resulting object can no longer be considered a remnant but some type of 
    stellar stream (proper motions of the member stars remain similar 
    although no overdensity can be recorded). For the Solar Neighbourhood, 
    the stellar density of the system should be greater than 0.044 $M_{\odot}$ 
    pc$^{-3}$. Moreover, the density of a star cluster in the Solar 
    Neighbourhood must be greater than about 0.08 $M_{\odot}$ pc$^{-3}$ in 
    order to be stable against tidal disruption. Taking these constraints into 
    consideration, the smallest OCRs from poorly populated clusters ($N \leq 
    100$) should contain more than 40\% of their initial population in order 
    to have a stellar density larger than the surrounding field. This value is 
    close to the one quoted in Bica et al. (2001) but the situation appears to 
    change dramatically for richer clusters. OCRs from clusters having 
    $N$ in the range 200-500 should retain 10-20\% of their initial members. 
    For $N$ in the range 750-1,000, they should keep about 7\% of their initial 
    population. Densely populated clusters ($N$ = 10,\,000) generate OCRs 
    containing about 0.1\% of their initial population, but with stellar 
    density in the range 0.3-15 $M_{\odot}$ pc$^{-3}$. OCRs from rich
    clusters are very different from those resulting from the evolution of
    small clusters. The surviving stellar system is the outcome of a long-term
    contest for dynamical stability and, typically, only highly stable 
    hierarchical multiple systems are present there. They are the result of
    Gyrs of dynamical evolution.

    Observationally, Bica et al. (2001) selected 35 groups that stand up 
    against the mean Galactic stellar fields, and are located relatively high 
    onto the Galactic plane.  Regrettably, the close scrutiny of the 
    kinematics in some of them (Villanova et al. 2004a; Carraro et al. 2005;
    Carraro 2006) has shown that the contrast criterion is necessary but not 
    sufficient to detect a remnant. In fact, when dealing with the remnant 
    of a physical group, one expects that the remains itself exhibits all the 
    features of a physical group. Following Platais et al. (1998), we define 
    a physical group (at odds with a random sample of field stars) as an 
    ensemble of stars which (1) occupy a limited volume of space, 
    (2) individually share a common space velocity, and (3) individually 
    share the same age and chemical composition, producing distinctive 
    features in the H-R diagram.

    In an attempt to provide tight observational constraints for numerical 
    models, we have studied in detail the star group NGC~1901, which appears
    to be a promising open cluster remnant candidate. In this paper, we 
    present new CCD UBVI photometry and 3 epoch spectroscopy, which we 
    combine with proper motions from the UCAC2 catalog (Zacharias et al. 2004), 
    with the aim to clarify the nature and dynamical status of this cluster.
    In Sect. 2 we provide a historical overview of the various analyses
    carried out previously for this object. Observations, data reduction, and
    overall results are presented in Sect. 3. In Sect. 4 we analyse proper
    motions and in Sect. 5 we discuss cluster members. In Sect. 6 we derive
    cluster fundamental parameters. Section 7 is devoted to identify possible 
    fainter members. The current dynamical status is analysed in Sect. 8 and 
    the connection with the Hyades stream is considered in Sect. 9. In Sect. 10 
    we draw some conclusions and suggest further lines of research.
    \begin{figure}
     \centering
      \includegraphics[angle=-90,width=9cm]{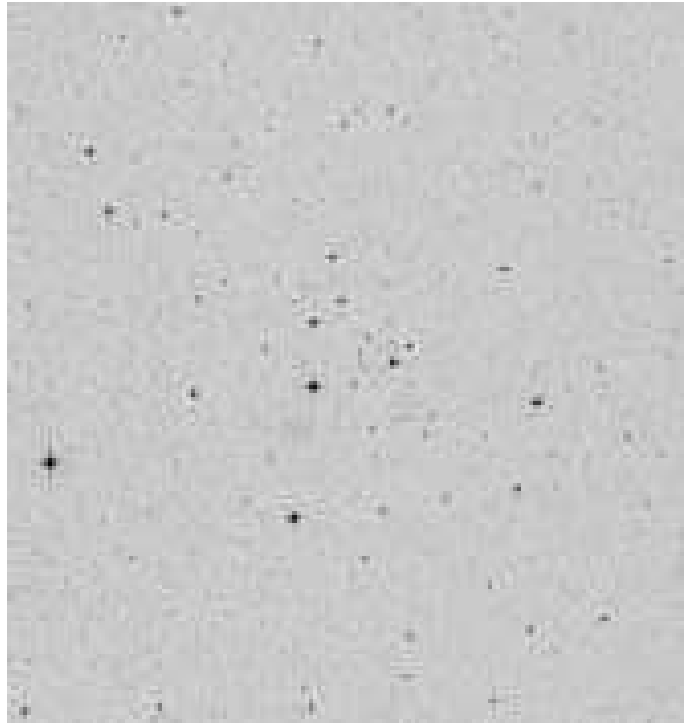}
      \caption{DSS image of the area covered by the present study.
               North is up, east to the left, and the image is 
               20 arcmin on a side.}%
    \end{figure}
%

%
%
\section{Historical overview}
    The star group NGC~1901 was discovered in 1836 by J. F. W. Herschel
    but first noticed as a possible physical group by Bok \& Bok (1960) 
    while re-defining the van Wijk (1952) photometric sequence in front of 
    the Large Magellanic Cloud (LMC). Bok \& Bok (1960) emphasize that the 
    region of the van Wijk sequence possibly represents a loose association 
    of stars projected against the Large Cloud, since in this direction 
    ($\alpha=05^{\rm h}~18^{\rm m}~11^{\rm s}$, $\delta=-68^{\circ} 
    27^{\prime}$, $l=279^{\circ}.03$, $b=-33^{\circ}.60$, J2000) the LMC is 
    not very dense.

    The photoelectric BV photometry collected by Sanduleak \& Philip (1968, 
    hereinafter SP68) allowed indeed to find out for the first  time 
    that the stars in the association compose a sequence in the Color Magnitude 
    Diagram (CMD), compatible with a group of stars at 330 pc from the Sun.
    This result was noticed to be reinforced by the preliminary objective-prism
    radial velocities taken a few years before by Fehrenbach (1965), which 
    showed how a few stars did have radial velocities smaller than the expected
    value (larger than 275 km s$^{-1}$) for objects belonging to the LMC. For 
    some of these stars, SP68 provided proper motions from a variety of 
    sources, showing that a substantial fraction of them did share common 
    motion. A more detailed proper motion analysis was performed the same year 
    by Murray et al. (1969), which confirmed the reality of the group but put 
    it a 480 pc from the Sun, and called the attention to a possible dynamical 
    relation with the Hyades, due to the very similar space motion components 
    of the two groups. This argument has been thoroughly investigated by 
    Eggen (1996), who found that the two systems are part of the Hyades 
    stream. He found two separate kinematic groups that he called 
    superclusters within the Hyades stream: the Hyades and the NGC~1901 
    superclusters. Using the UVW system defined in Sect. 9 the heliocentric
    space motion of the Hyades group is (U, V, W) = (-40.4, -16.0, -3.0) km/s
    and the equivalent result for the NGC~1901 group is   
    (U, V, W) = (-26.4, -10.4, -1.5) km/s. Although these results strongly
    suggest that two different kinematic structures are being observed,
    no discussion of errors or dispersions is given in Eggen's paper.
    Using samples observed by the Hipparcos satellite, Dehnen (1998) confirmed 
    the present of two distinct moving groups in the area. The Hyades group 
    was characterized by (U, V, W) = (-40, -20, 0) km/s and for the NGC~1901 
    group he found (U, V, W) = (-25, -10, -15) km/s. Chereul et al. (1999)
    give (U, V, W) = (-32.9, -14.5, -5.6) km/s with velocity dispersions
    ($\sigma_U$, $\sigma_V$, $\sigma_W$) = (6.6, 6.8, 6.5) km/s for the
    Hyades group although they found three different groups within the 
    supercluster. The NGC~1901 group is found at (U, V, W) = 
    (-26.1, -7.6, -0.7) km/s with velocity dispersions 
    ($\sigma_U$, $\sigma_V$, $\sigma_W$) = (2.1, 3.7, 3.2) km/s. Again,
    both kinematic structures appear well separated.

    \noindent
    Finally, a new investigation has been recently published by Pavani et al. 
    (2001, hereinafter P01). They extended the photometry in a small area 
    down to $V$ = 16. However, the poor quality of the photometry and the 
    lack of any new kinematic information prevented them from adding 
    new pieces of information.

    \section{Observational material: Photometry}
       $U,B,V,$ and $I$ images centered on NGC~1901  were obtained at the 
       Cerro Tololo Inter-American Observatory 1.0 m telescope, which is 
       operated by the SMARTS\footnote{http://www.astro.yale.edu/smarts/} 
       consortium. The telescope is equipped with a new 4k$\times$4k CCD 
       camera having a pixel scale of 0$^{\prime\prime}$.289/pixel which 
       allows one to cover a field of $20^{\prime} \times 20^{\prime}$. 
       Observations were carried out on November 29, 2005. Three Landolt 
       (1992) areas (TPhoenix, Rubin~149, and PG~0231+006) were also 
       observed to tie the instrumental magnitudes to the standard system. 
       The night was photometric with an average seeing of 1.1 arcsec. Data 
       have been reduced using IRAF\footnote{IRAF is distributed by NOAO, which 
       is operated by AURA under cooperative agreement with the NSF.}
       packages CCDRED, DAOPHOT, and PHOTCAL. Photometry was done employing
       the point spread function (PSF) method (Stetson 1987). The covered
       area is shown in Fig.~1, while Table~1  contains the
       observational log-book.\\
       The calibration equations read: 

\begin{table}
\fontsize{8} {10pt}\selectfont
\caption{Journal of photometric observations of NGC~1901 and standard star
         fields together with calibration coefficients (November 29, 2005).}
\begin{tabular}{ccccccc}
\hline
\multicolumn{1}{c}{Field}         &
\multicolumn{1}{c}{Filter}        &
\multicolumn{1}{c}{Exposure time} &
\multicolumn{1}{c}{Seeing}        &
\multicolumn{1}{c}{Airmass}       \\
 &  \multicolumn{3}{c}{[sec.]} & [$\prime\prime$] & \\
\hline
NGC 1901        & U & 1200,60,5 & 1.1 & 1.150-1.280 \\
                & B &  900,30,3 & 1.0 & 1.150-1.280 \\
                & V &  600,30,1 & 1.0 & 1.150-1.280 \\
                & I &  600,30,1 & 1.0 & 1.150-1.280 \\
\hline
TPhoenix        & U & 180,200   & 1.0 & 1.024,1.444 \\
                & B &  90,120   & 1.1 & 1.023,1.447 \\
                & V &  20,30    & 1.1 & 1.024,1.450 \\
                & I &  40,40    & 1.1 & 1.022,1.452 \\
\hline
PG 0231+006     & U & 200,240   & 1.1 & 1.291,1.801 \\
                & B &  60,90    & 1.1 & 1.293,1.807 \\
                & V &  40,40    & 1.1 & 1.296,1.809 \\
                & I &  40,30    & 1.1 & 1.294,1.810 \\
\hline
Rubin~149       & U & 180,240   & 1.0 & 1.311,1.651 \\
                & B &  90,120   & 1.1 & 1.316,1.649 \\
                & V &  30,40    & 1.0 & 1.318,1.647 \\
                & I &  40,40    & 1.0 & 1.313,1.643 \\
\hline
\hline
Calibration     & \multicolumn {3}{l}{$u_1 = +3.285 \pm 0.004$} \\
coefficients    & \multicolumn {3}{l}{$u_2 = +0.032 \pm 0.006$} \\
                & \multicolumn {3}{l}{$u_3 = +0.46$}            \\
                & \multicolumn {3}{l}{$b_1 = +2.188 \pm 0.004$} \\
                & \multicolumn {3}{l}{$b_2 = -0.160 \pm 0.006$} \\
                & \multicolumn {3}{l}{$b_3 = +0.27$}            \\
                & \multicolumn {3}{l}{$v_{1bv} = +2.188 \pm 0.014$} & \multicolumn {3}{l}{$i_1 = +2.789 \pm 0.044$} \\
                & \multicolumn {3}{l}{$v_{2bv} = +0.017 \pm 0.018$} & \multicolumn {3}{l}{$i_2 = +0.021 \pm 0.043$} \\
                & \multicolumn {3}{l}{$v_3 = +0.12$}                & \multicolumn {3}{l}{$i_3 = +0.06$}            \\
                & \multicolumn {3}{l}{$v_{1vi} = +2.188 \pm 0.016$} & \\
                & \multicolumn {3}{l}{$v_{2vi} = +0.013 \pm 0.016$} & \\
\hline
\end{tabular}
\end{table}

          \begin{center}
           \begin{tabular}{lc}
            $u = U + u_1 + u_2 (U-B) + u_3 X$         & (1) \\
            $b = B + b_1 + b_2 (B-V) + b_3 X$         & (2) \\
            $v = V + v_{1bv} + v_{2bv} (B-V) + v_3 X$ & (3) \\
            $v = V + v_{1vi} + v_{2vi} (V-I) + v_3 X$ & (4) \\
            $i = I + i_1 + i_2 (V-I) + i_3 X$         & (5), \\
           \end{tabular}
          \end{center}

         \noindent
         where $UBVI$ are standard magnitudes, $ubvi$ are the instrumental 
         ones, $X$ is the airmass and the derived coefficients are presented 
         at the bottom of Table~1. As for $V$ magnitudes, when $B$ magnitude 
         was available, we used expression (3) to compute them, elsewhere 
         expression (4) was used. The standard stars in these fields provide 
         a wide color coverage, with
         $-1.217\leq (U-B) \leq 2.233$,$-0.298\leq (U-B) \leq 1.999$,
         and $-0.361\leq (V-I) \leq 2.268$. Aperture correction was estimated 
         in a sample of bright stars, and then applied to all the stars. It 
         amounted at 0.315, 0.300, 0.280 and 0.280 mag for the $U$, $B$, $V$ 
         and $I$ filters, respectively.

    \noindent
    We now compare our photometry with previous investigations. We restrict 
    this comparison to the photoelectric photometry of SP68 and the modern
    CCD study by P01. We have 12 stars in common with SP68, and the comparison
    in the sense this study minus SP68 reads:

   \begin{equation}
     \Delta (B - V) = 0.024 \pm 0.044
    \end{equation}

    \begin{equation}
     \Delta V = 0.012 \pm 0.056
    \end{equation}

     \noindent
    In general, this comparison is very good and it is shown in 
    Fig.~2.\\ 
    Now we turn to the only previous CCD investigation by P01. We 
    have 30 stars in common and the comparison in the sense this study minus 
    P01 reads:

    \begin{equation}
     \Delta (B - V) = 1.972 \pm 4.075
    \end{equation}

    \begin{equation}
     \Delta V = -0.267 \pm 0.619
    \end{equation}

    \noindent
    In Fig.~3 only a sub-sample of these 30 stars is shown.\\
    Clearly some errors occurred in P01 photometry. P01 
    photometry compares well with SP68 and the present study down to $V$ 
    $\sim$ 11.5, then it starts deviating   increasingly with respect to 
    the present study. We notice that P01 failed to compare their photometry
    with two more stars they have in common with SP68 (GSC091200626 and 
    GSC092100834). While the present study compares very well with SP68 also 
    for these two stars, P01 shows significant deviations (up to 0.5 mag in 
    $V$) for these two stars. The comparison with P01 is shown in
    Fig.~3. Another strange point is that P01 assign two entries
    to TYC-2883 and GCS~0916200883 (actually the same
    star, HD~269310), but provide for it two
    very discrepant values of the magnitude and color and different
    suggestions about the membership.\\
%
%
    \begin{figure}
     \centering
      \includegraphics[width=9cm]{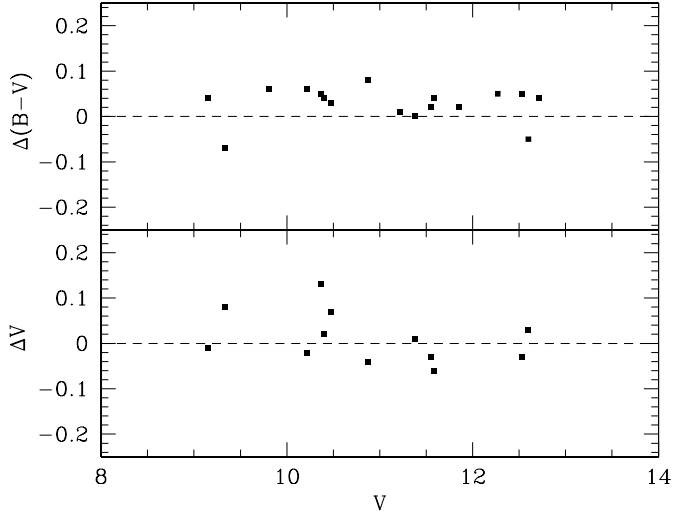}
      \caption{Comparison with SP68 photoelectric photometry.
               The comparison is in the sense this study minus SP68}%
    \end{figure}
%

%
    \begin{figure}
     \centering
      \includegraphics[width=9cm]{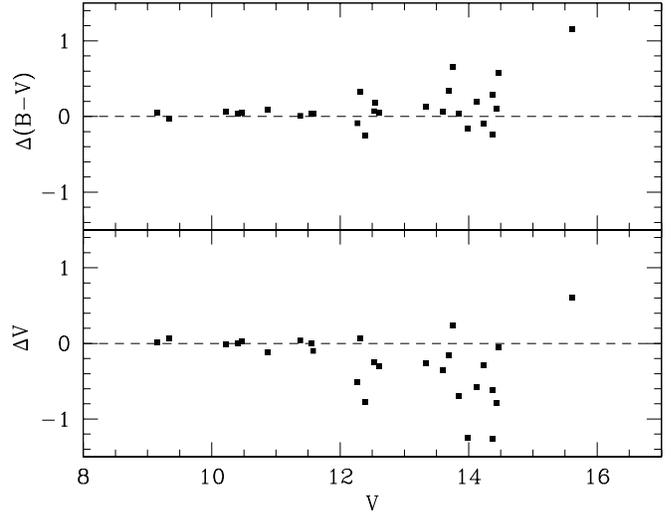}
      \caption{Comparison with P01 CCD photometry. The comparison
               is in the sense this study minus P01. In this figure
               the two most deviating stars are not shown.}%
    \end{figure}
    
    \noindent
    The CMD of all the stars measured in the field of NGC~1901 (see Fig.~1)
    is shown in Fig.~4. The present photometry goes 7 mag deeper than P01
    and provides a larger wavelength baseline (from $U$ to $I$). The CMD 
    in Fig.~4 is in the $V$ vs $B - V$ plane, to compare with previous 
    studies. The objective is to extend NGC~1901 sequence as deep as possible.
    The main problem is, of course, the strong contamination from LMC stars.
    The LMC is dramatically visible in this CMD, with its young star
    Main Sequence (MS), intermediate-age population red clump, Asymptotic 
    Giant Branch (AGB), and the older stars Red Giant Branch (RGB).

    \begin{figure}
     \centering
      \includegraphics[width=\columnwidth]{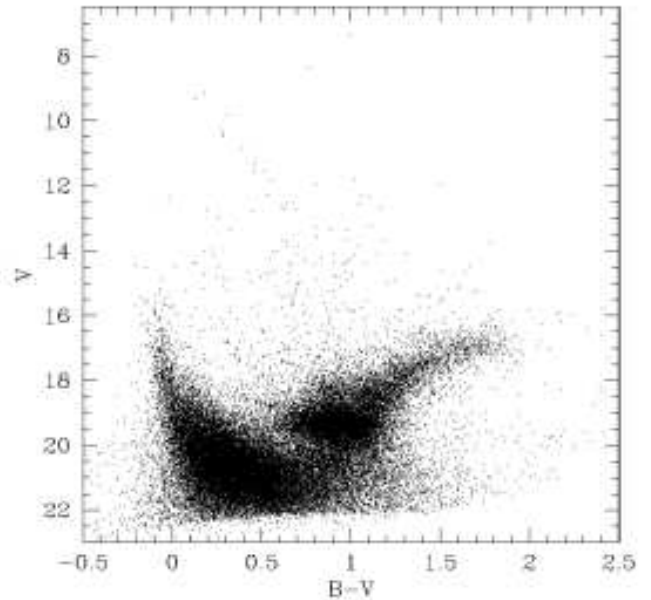}
      \caption{BV CMD for all the stars detected in the present study
	having $\sigma_B$ and $\sigma_V$ smaller than 0.1 mag.}%
    \end{figure}


   \begin{figure}
    \centering
     \includegraphics[width=9cm]{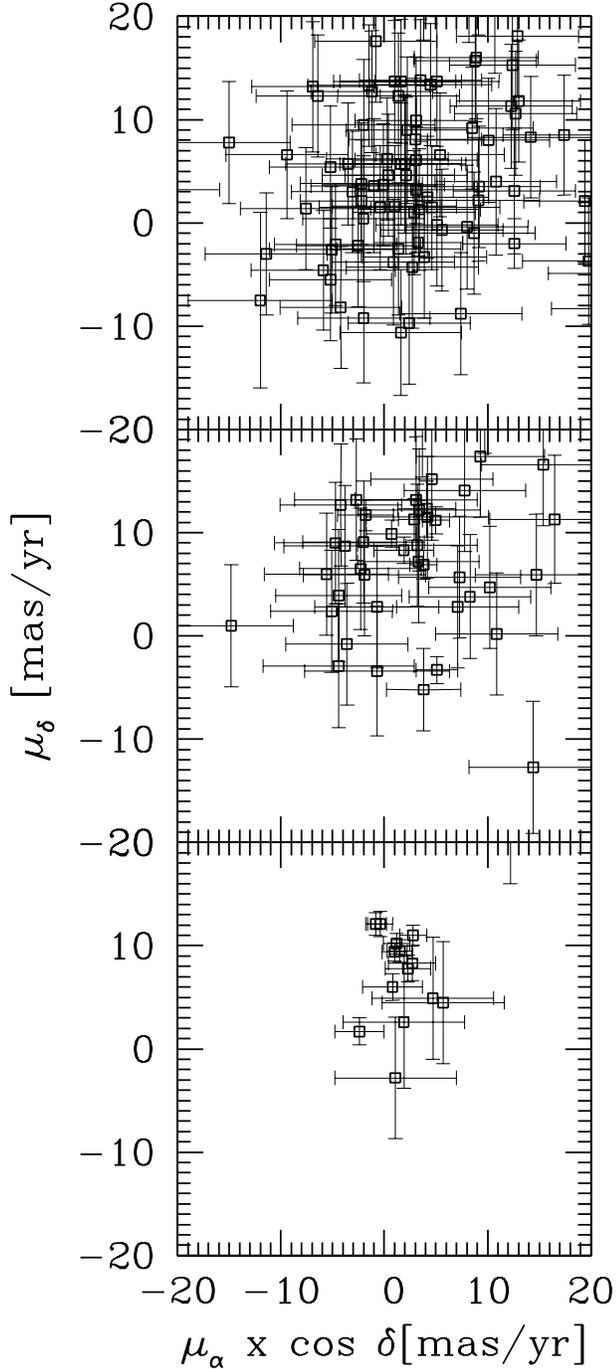}
     \caption{Proper motion analysis. {\bf Upper panel}:
     stars having $12 \leq K \leq 14$. {\bf Middle panel}: stars having 
     $10 \leq K \leq
     12$. {\bf Lower panel}: stars brighter than K = 10.0. These stars
     are used to search for a preliminary member list. }%
   \end{figure}
%

\section{Proper motions}
   We have extracted the proper motion components in a 20 squared arcmin field 
   around NGC~1901 from the UCAC2 catalogue (Zacharias et al. 2004),
   and 
   constructed 3 vector point diagrams as a function of K
   magnitude (UCAC2 is in fact cross-correlated with 2MASS). The results are
   shown in Fig.~5, where the lower panel is restricted to stars
   having K $\leq$ 10, the middle panel to stars in the range $10 \leq
   K \leq 12$, and the upper panel to stars in the range $12 \leq K
   \leq 14$. 
   The  contamination from background stars is very high in the upper and middle
   panel. For this reason, we consider only the stars brighter than K
   = 10 (lower panel) to have an estimate of the cluster mean tangential motion and
   derive a first list of member candidates.
   In such way we extracted 35 stars
   from the catalog, from which we derived :

   \begin{equation}
    \mu_{\alpha} = 1.7 \pm 1.3 ~[ms/yr]
   \end{equation}

   \begin{equation}
    \mu_{\delta} = 12.3 \pm 2.9 ~ [mas/yr]
   \end{equation}

   \noindent
   Clearly, stars in the lower panel seem to visually clump around these values.
   Starting from here, we are going to consider proper motion probable 
   members all the stars having proper motion component compatible within 
   2$\sigma$ of these values.

    \section{Observational material: Spectroscopy}
   To obtain more solid information about the membership and dynamical status 
   of NGC~1901 we carried out a multi-epoch spectroscopic campaign.
   With the aim to measure radial velocity and detect unresolved binaries, 
   spectroscopy has been obtained during three different runs in 2002, 2003
   and 2005.

        \subsection{La Silla data}
            The first epoch of radial velocity observations was already
            described in Villanova (2003) and 
            Villanova et al. (2004b) and were carried out on the night of 2002 
            December 10 at the La Silla Observatory (ESO, Chile) under photometric conditions 
            and a typical seeing of 1 arcsec. The EMMI spectrograph on the 
            NTT 3.5-m telescope was used with a 1.0'' slit to provide a 
            spectral resolution R = 33000 in the wavelength range 
            3800-8600 \AA \ on the two 2048$\times$4096 CCDs of the mosaic 
            detector. For the wavelength calibration, a single spectrum of 
            a thorium-argon lamp was secured at the end of the night because 
            of the stability of the instrument. The typical
            signal-to-noise of the spectra range from 70 to 170.
 
        \subsection{CASLEO data}
            This observational dataset consists of 4 high resolution spectra 
            for radial velocities determination and 2 low resolution spectra 
            for MK classification.
            The 4 high resolution spectra have been obtained in
            December 11-15, 2003 with the 
            Echelle REOSC Cassegrain spectrograph attached to the 2.15m 
            reflector at the Complejo Astronomico El Leoncito (CASLEO 
            \footnote{CASLEO is operated under agreement between CONICET, 
            SECYT, and the National Universities of Cordoba, La Plata and 
            San Juan, Argentina.}), using a TEK 1024$\times$1024 pixel CCD as 
            detector. The spectra cover an approximate wavelength range from 
            3600 to 6000 $\AA$, at a reciprocal dispersion of $\sim$ 0.15 
            $\AA$ per pixel at 4000 $\AA$.
            Comparison arc images were observed at the same telescope 
            position as the stellar images immediately before and after the 
            stellar exposures. Spectra were all reduced using available IRAF 
            routines.
            The radial velocities were determined on spectra normalized to 
            the continuum, fitting Gaussian profiles to the observed line 
            using IRAF routines, measured the position of the line or the two 
            components presents in the echelle spectra. The 2 low resolution 
            spectra observed from CASLEO, in December 8, 2004 with the same 
            REOSC spectrograph as echelle but in its simple dispersion mode, 
            provides on the TEK 1024$\times$1024 a sampling of 1.64 $\AA$ per 
            pixel, covering a wavelength range from 3900 to 5500 $\AA$, over 
            the classical range for MK classification. All spectra were 
            reduced using standard IRAF routines.

        \subsection{Las Campanas data}  
            The third epoch radial velocity observations were carried out 
            on the nights of 2005 October 18-22 at the du Pont 2.5-m 
            telescope at Las Campanas Observatory under variable photometric 
            conditions 
            and a typical seeing around $1-2^{\prime\prime}$. 
            The echelle spectrograph was used 
            with a 0.7'' slit to provide a spectral resolution R = 64000 in 
            the wavelength range 3600-10400 \AA \ on the 2048$\times$2048 CCD. 
            Images were reduced using IRAF including bias subtraction, 
            flat-field correction, extraction of spectral orders, wavelength 
            calibration, sky subtraction, and spectral rectification. The 
            single orders were merged into a single spectrum. 
            The typical
            signal-to-noise of the spectra ranges from 50 to 80.

            \noindent
            The radial velocities of the target stars in the La Silla
            and Las Campanas runs were measured using the 
            IRAF {\tt fxcor} task, which cross-correlates the object spectrum 
            with the template. The peak of the cross-correlation was fitted 
            with a Gaussian curve after rejecting the spectral regions 
            contaminated by telluric lines ($\lambda > 6850 \AA$). As templates 
            we calculated a set of spectra covering all the T$_{eff}$ range 
            (4500-9000 K) of our stars using SPECTRUM, the Local
            Thermodynamical Equilibrium (LTE) spectral 
            synthesis program freely distributed by Richard O'Gray 
            \footnote{ www.phys.appstate.edu/spectrum/spectrum.html}.\\

            \noindent
            The obtained radial velocities for the three epochs are reported
            in Table~2 together with their errors, and plotted in
            Fig.~6.  

%
%
     \begin{figure}
      \centering
       \includegraphics[width=9cm]{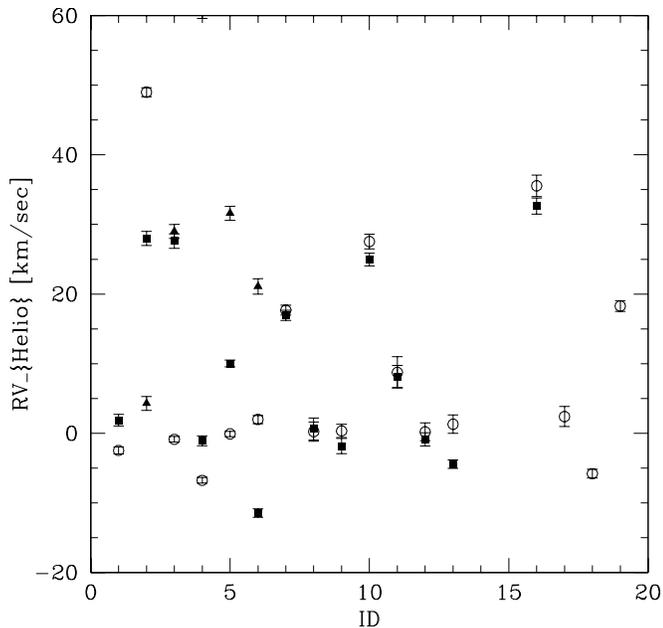}
       \caption{Radial velocity distribution. Open circles indicate
         RVs from the Las Campanas run, filled squares indicate
         RVs from the La Silla run, and filled triangles the Casleo run.}%
     \end{figure}

%
     \begin{figure}
      \centering
       \includegraphics[width=9cm]{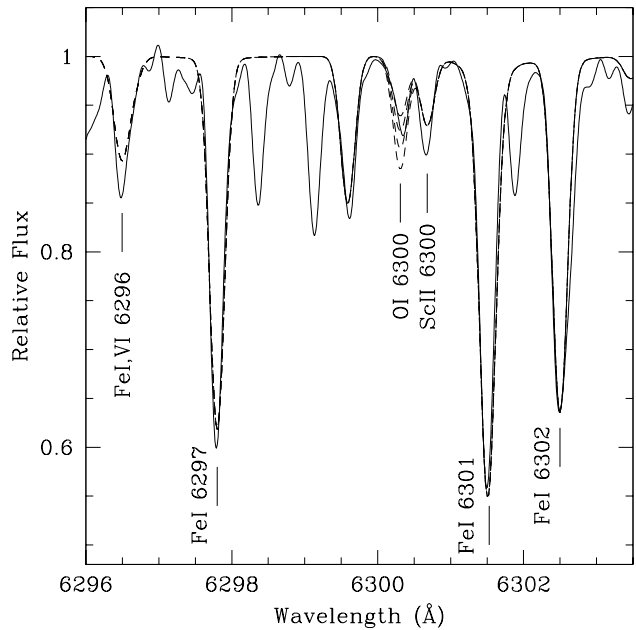}
       \caption{Oxygen synthesis around 6300~\AA \ region. The oxygen
       abundance covers the range from -0.2 dex (upper dashed line at
       6300 ~\AA) to +0.1 dex (lower dashed line at 6300 ~\AA).}%
     \end{figure}

\begin{table*}
 \fontsize{8} {10pt}\selectfont
 \tabcolsep 0.10truecm
 \caption{Basic data of suspected NGC~1901 members. In the last column
 M indicates membership to NGC~1901, NM no membership.}
 \begin{tabular}{ccccccccccccccc}
 \hline
 \multicolumn{1}{c}{ID}         &
 \multicolumn{1}{c}{Designation}        &
 \multicolumn{1}{c}{$\alpha$} &
 \multicolumn{1}{c}{$\delta$}        &
 \multicolumn{1}{c}{$U$}       &
 \multicolumn{1}{c}{$B$}       &
 \multicolumn{1}{c}{$V$}       &
 \multicolumn{1}{c}{$I$}       &
 \multicolumn{1}{c}{$\mu_{\alpha}$}       &
 \multicolumn{1}{c}{$\mu_{\delta}$}       &
 \multicolumn{1}{c}{$RV Dic02$} &
 \multicolumn{1}{c}{$RV Dic03$} &
 \multicolumn{1}{c}{$RV Oct05$} &
 \multicolumn{1}{c}{$Spectral Type$}& 
 \multicolumn{1}{c}{Membership}\\
 \hline
 & & hh:mm:sec & $o$:$\prime$:$\prime\prime$& mag & mag & mag & mag & mas/yr & ms/yr & km/sec & km/sec & km/sec & &\\
\hline
  2&       HD35294& 05:18:03.258&  -68:27:56.69&      9.63&    9.15&   8.38&   7.55&    1.0$\pm$1.2&       9.4$\pm$0.9&               & 4.3$\pm$1.0  &  2.4$\pm$1.4 &G2IV&M\\
  3&       HD35183& 05:17:23.034&  -68:28:19.12&      9.51&    9.33&   9.15&   8.94&    1.5$\pm$1.2&       9.5$\pm$1.1&  1.8$\pm$0.8  & 29.0$\pm$1.0 &-2.5$\pm$0.5 &A3V&M\\
  4&       HD35293& 05:18:02.089&  -68:21:19.48&      9.66&    9.46&   9.33&   9.12&   -0.8$\pm$1.0&      12.1$\pm$1.1& 27.9$\pm$1.0  & 60.6$\pm$1.0 & 48.9$\pm$0.7 &A1&M\\ 
  5&       HD35462& 05:19:10.754&  -68:34:51.72&     10.28&   10.13&   9.81&   9.49&    1.2$\pm$1.3&      10.2$\pm$1.0&               & 31.6$\pm$1.0 & 18.3$\pm$0.8 &A2V&M\\
  6&      HD269338& 05:18:52.918&  -68:34:13.35&     10.76&   10.65&  10.37&  10.08&   -0.4$\pm$1.2&      12.1$\pm$1.2&        & 21.0$\pm$1.1  & -5.8$\pm$0.6 &A5&M\\
  7&      HD269319& 05:18:22.479&  -68:28:01.63&     10.64&   10.49&  10.22&   9.97&    2.8$\pm$1.3&      11.0$\pm$1.0& 27.6$\pm$1.0  &       & -0.9$\pm$0.4 &A5&M\\
  8&      HD269312& 05:18:11.983&  -68:25:36.34&     10.82&   10.68&  10.40&  10.11&    2.3$\pm$2.2&       7.8$\pm$1.3& -1.1$\pm$0.7  &       & -6.7$\pm$0.4 &&M\\
  9&      HD269310& 05:17:59.168&  -68:31:27.72&     10.92&   10.75&  10.46&  10.14&    2.7$\pm$2.3&       8.3$\pm$1.7& 10.0$\pm$0.4  &       & -0.1$\pm$0.4 &A7V&M\\
 10&      HD269301& 05:17:35.815&  -68:21:47.16&     11.31&   11.26&  10.87&  10.56&    5.0$\pm$1.3&      11.2$\pm$1.3& -11.4$\pm$0.6 &       &  2.0$\pm$0.6 &A9V&M\\
 12&      HD269315& 05:18:16.683&  -68:25:10.44&     11.84&   11.83&  11.38&  10.86&    3.8$\pm$1.3&       6.9$\pm$1.4& 16.9$\pm$0.7  &       & 17.7$\pm$0.7 &F5V&NM\\
 13&      HD269324& 05:18:29.529&  -68:27:13.81&     12.11&   12.07&  11.55&  10.95&    2.9$\pm$1.5&      11.3$\pm$2.0&  0.6$\pm$1.6  &       & 0.2$\pm$1.4  &F6V&M\\
 14&      HD269334& 05:18:42.738&  -68:27:32.76&     12.04&   12.03&  11.58&  11.09&    1.9$\pm$2.9&       8.3$\pm$1.3&  -1.8$\pm$1.1 &       & 0.3$\pm$0.9  &F3V&M\\
 17&  GSC916200464& 05:18:27.560&  -68:31:28.36&     11.71&   12.37&  12.39&  12.35&   12.6$\pm$5.0&      -2.0$\pm$2.4& 268.6$\pm$1.6  &       &       &G5V&NM\\
 20&  GSC916200834& 05:19:05.214&  -68:30:45.75&     12.74&   12.80&  12.27&  11.67&   -1.8$\pm$2.7&      11.7$\pm$1.5& -0.9$\pm$0.9  &       &  0.2$\pm$1.3 &F8V&M\\
 23&  GSC916200682& 05:18:41.691&  -68:22:29.24&     14.07&   13.36&  12.31&  11.38&    1.1$\pm$5.9&      -2.8$\pm$5.9& 8.2$\pm$1.6   &       & 8.7$\pm$2.2 &K0&NM\\
 25&  GSC916201005& 05:17:26.939&  -68:25:41.23&     13.32&   13.16&  12.53&  11.79&    1.6$\pm$5.9&      30.4$\pm$5.9& 24.9$\pm$0.9  &       & 27.5$\pm$1.0 &F7&NM\\
 27&  GSC916200626& 05:18:19.173&  -68:26:25.15&     13.21&   13.21&  12.60&  11.97&    0.7$\pm$2.4&       9.9$\pm$1.3& -3.4$\pm$0.6  &       &   1.3$\pm$1.3&F8V&M\\
 36& MACS0517684015& 05:17:40.773&  -68:29:04.73&     14.52&   14.22&  13.34&  12.35&   10.3$\pm$5.9&      32.6$\pm$5.9& 32.6$\pm$1.1  &       & 35.5$\pm$1.5&&NM\\
 40&   UCAC2-139  & 05:18:25.564&  -68:17:19.25&     14.48&   14.37&  13.74&  12.96&   -2.2$\pm$5.9&       3.8$\pm$5.9& 295.9$\pm$2.4 &       & 298.8$\pm$3.5&&NM\\
 41& GSC0916200216& 05:17:51.066&  -68:24:34.62&     17.43&   15.54&  13.76&  12.15&    5.7$\pm$5.9&       4.5$\pm$5.9&  30.1$\pm$0.1 &       &  32.2$\pm$0.2&&NM\\
 \hline
 \end{tabular}
\end{table*}

%
%
%
   \begin{figure}
    \centering
     \includegraphics[width=9cm]{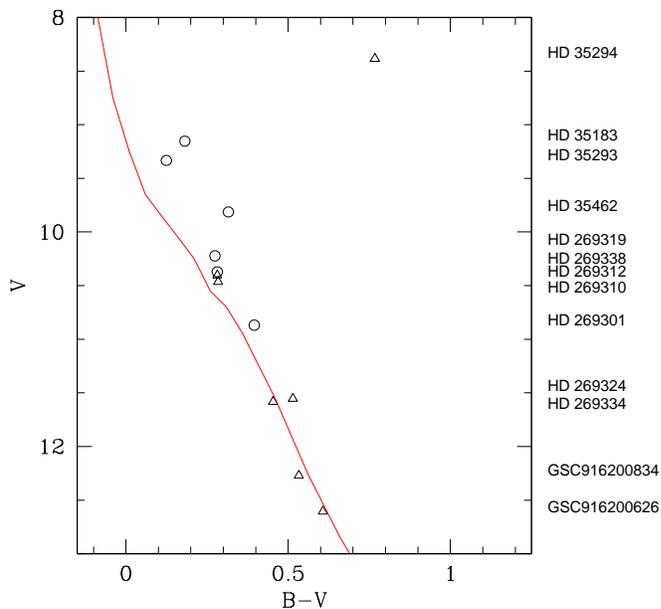}
     \caption{Color Magnitude Diagram for cluster members. 
              Circles indicate binaries, while triangles indicate
              single stars.}%
   \end{figure}
%

%
   \begin{figure}
    \centering
     \includegraphics[width=9cm]{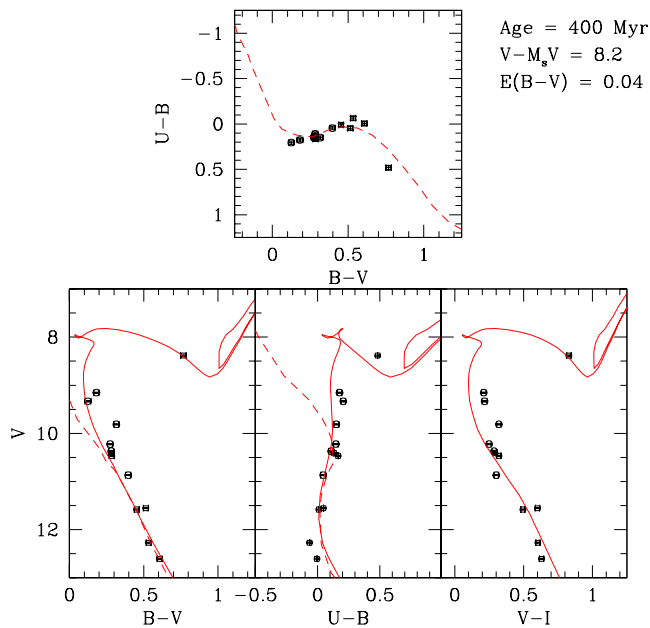}
     \caption{Color Magnitudes and Two Color Diagrams: looking for NGC~1901 
              fundamental parameters. The upper panel shows the
              TCD, while the three bottom panels shows the CMDs for
              different color combinations.
              The solid line in all the panels
               is the same isochrone shifted to fit the stars distribution, while
              the dashed line is the Schmidt-Kaler (1982) empirical ZAMS.}%
   \end{figure}
\section{Cluster members}  
   In this Section we make use of the 
   stellar radial velocities, proper motions and positions in the various 
   photometric diagrams to deem out interlopers and pick up cluster
   members. 
   We consider here at first only the stars which 
   have radial velocity, proper motion and multicolor photometry. With 
   these stars we try to define the cluster MS, in order to be able
   later on to detect fainter photometric members along the extension
   of this MS. In the following we discuss the group membership on
   a star by star basis.\\

   \noindent
   {\bf HD35294.}
   This is a giant star of G2IV spectral type. Its radial velocity suggests 
   that it is a single star and a member of NGC~1901. Fehrenbach \& Duflot 
   (1974) found similar values for the radial velocity of this star, reassuring
   its single nature. Proper motions are also compatible with membership.\\

   \noindent
   {\bf HD35183.}
   This is clearly a binary star, but proper motions do suggest membership.\\

   \noindent
   {\bf HD35293.}
   This is another clear binary star, and again proper motions do suggest
   membership. \\

   \noindent
   {\bf HD35462.} 
   Proper motions suggest that it is a member, but radial velocities
   seem to contradict this hypothesis. 
   However, if we combine together our radial velocity
   measurements with older data (11 Km s$^{-1}$, Barbier-Brossat et al. 1994)
   it emerges that this is a binary star, therefore reinforcing the idea
   that it is a member of the group.\\

   \noindent
   {\bf HD269338.}
   This star is a binary, and proper motions and radial velocity are consistent 
   with it being a cluster member.\\

   \noindent
   {\bf HD269319.}
   This star is a binary as well, and proper motions and radial velocity are 
   consistent with it being a cluster member.\\

   \noindent
   {\bf HD269312.}Proper motions and radial velocity suggest this is a cluster 
   member. This star has recently been studied by Cherix et
   al. (2006), who found that it is a $\delta$ Scuti variable candidate.\\

   \noindent
   {\bf HD269310.}
   This star is a binary, and proper motion and radial velocity
   identify it 
   as a cluster member. Binarity is confirmed by Barbier-Brossat 
   et al. (1994), who report values of -16 and 10 km/sec for this star.\\

   \noindent
   {\bf HD269301.} 
   This star is a binary, and proper motion and radial velocity indicate 
   that it is a cluster member.\\

   \noindent
   {\bf HD269315.} 
   We propose this star is not a member. Radial velocity is significantly 
   larger than the cluster average and proper motions also deviate from
   the mean tangential motion.\\

   \noindent
   {\bf HD269324.}
   Both radial velocity and proper motion indicate that this star is a 
   member of NGC ~1901.\\

   \noindent
   {\bf HD269334.} 
   As in the previous case, both radial velocity and proper motion 
   are suggesting that this star is a member of  NGC~1901.\\

   \noindent
   {\bf GSC0916200464.} 
   The high radial velocity, blue color and proper motion readily indicate
   that this is a LMC star.\\

   \noindent
   {\bf GSC0916200834.} 
   This star is a clear member considering its proper motion and radial 
   velocity.\\

   \noindent
   {\bf GSC0916200682.} 
   P01 propose this is a member. However it looks a clear proper motion 
   non-member, and the radial velocity confirms this conclusion.\\

   \noindent
   {\bf GSC0916201005.} 
   Murray et al. (1969) already suggested that this is a non-member, although 
   P01 propose it as a photometric member. Our radial velocity and UCAC~2 
   proper motions confirm Murray et al. results.\\

   \noindent
   {\bf GSC0916200626.}
   This star was found to be a proper motion non-member by Murray et al. 
   (1969), but P01 include it among the list of probable cluster members. 
   According to the UCAC2 proper motions, this is a likely member. Radial 
   velocity is compatible with membership, although it may be a binary.\\

   \noindent
   {\bf MACS0517684015.}
   Both the radial velocity and proper motion of this binary demonstrate that 
   it is a non-member.\\

   \noindent
   {\bf UCAC2-139.}
   The high radial velocity of this star suggests that it is a probable LMC
   member.\\

   \noindent
   {\bf GSC0916200216.}
   This is a single star for which both the radial velocity and the
   proper motion indicate it is an interloper.\\

    \section{NGC~1901 metal content}
        A chemical composition study of the cluster was performed by analyzing 
        the spectrum of the star HD 35294, which we have shown to be a
        member. Being the coldest stellar member, 
        it was the ideal target for this purpose. We used the equivalent widths 
        method and measured the equivalent width of the spectral lines for the 
        most important elements (O, Na, Mg, Si, Ca, Ti, Cr, Fe, Ni, Ba) using 
        the standard IRAF routine {\rm splot}. 
        Repeated measurements show a 
        typical error of about 5 $m\AA$ for the weakest lines. The LTE 
        abundance program MOOG (freely distributed by C. Sneden, University of 
        Texas, Austin) was used to determine the metal
        abundances. Model  
        atmospheres were interpolated from the grid of Kurucz (1992) models by 
        using the values of T$_{\rm eff}$ and log(g) determined as explained below. 
        Initial estimates of the atmospheric parameters (T$_{\rm eff}$, log(g), 
        and v$_t$) were obtained according to the spectral type of the star 
        (G2IV). Then, during the abundance analysis, T$_{\rm eff }$ and 
        v$_t$ were adjusted to remove trends in abundances vs. excitation 
        potential, and equivalent width for FeI lines 
        in order to obtain better estimates. Besides, log(g) was  adjusted
        in order to have the same abundance from {\rm FeI} and {\rm
        FeII} lines (ionization equilibrium).
        The final atmospheric parameters 
        for HD 35294 are $T_{eff}=5350\pm50 K$, $log(g)=3.2\pm0.1$, and 
        v$_t=1.08\pm0.05$ Km s$^{-1}$. The measured iron abundance is \rm{[Fe/H]} 
        = -0.08$\pm$0.02. Results for all the elements are reported in Table 3. 
        The O content was determined using both the forbidden line at 
        6300 $\AA$ and the permitted triplet at 7774 $\AA$. Because of 
        blending with other spectral lines, abundance from the forbidden line 
        was obtained comparing the observed spectrum with synthetic ones 
        calculated for different O contents (see Fig~7 and Table~3).
        Na, Mg, and O are well known to be affected by NLTE effects. For this 
        reason we applied abundance correction as prescribed by Gratton et al. 
        (1999). Using these corrections, the O abundance obtained from the 
        forbidden line (not affected by NLTE) agrees very well with the one 
        obtained from the O triplet (affected by NLTE). The chemical 
        composition shows a solar scaled mixture excepted for Ba, which is 
        super-solar of about 0.35 dex. Also Na and Mg, using the NLTE 
        corrections, are solar scaled as expected for a star cluster of this 
        metallicity.

\begin{table}
\fontsize{8} {10pt}\selectfont
\caption{NGC~1901 abundance analysis. NLTE correction is applied when necessary.}
\begin{tabular}{ccc}
\hline
\multicolumn{1}{c}{Element}         &
\multicolumn{1}{c}{LTE value}        &
\multicolumn{1}{c}{NLTE value} \\
\hline
 & dex & dex\\
\hline
\rm{[FeI/H]}    &-0.08$\pm$0.02&\\
\rm{[FeII/H]}    &-0.08$\pm$0.03&\\
\rm{[OI/H]7774}  &+0.05$\pm$0.09& -0.06$\pm$0.09\\
\rm{[OI/H]6300}  &-0.05$\pm$0.10&\\
\rm{[NaI/H]}     &-0.12$\pm$0.09& -0.02$\pm$0.09\\
\rm{[MgI/H]}     &-0.29$\pm$0.06& -0.13$\pm$0.06\\
\rm{[SiI/H]}     &-0.05$\pm$0.06&\\
\rm{[CaI/H]}     &-0.15$\pm$0.06&\\
\rm{[TiI/H]}     &-0.08$\pm$0.03&\\
\rm{[TiII/H]}    &-0.08$\pm$0.06&\\
\rm{[CrI/H]}     &-0.08$\pm$0.04&\\
\rm{[NiI/H]}     &-0.09$\pm$0.03&\\
\rm{[BaII/H]}    &+0.27$\pm$0.06&\\

\hline
\hline
\end{tabular}
\end{table}

\section{NGC~1901 reddening, distance and age}
   The CMD for the stars considered members of NGC~1901 is shown in Fig.~9
   for various colour combinations. In Fig.~9 the Two Color Diagram is also
   shown. In this figure we superpose an isochrone of 400 Myr, for the 
   exact metallicity, as derived in Section~7 for the star HD~35294. The 
   value $[Fe/H] = -0.08$ translates into $Z$ = 0.016 (Carraro et al. 1999). 
   In the upper panel however, we simply superpose an empirical Zero Age 
   Main Sequence (ZAMS) from Schmidt-Kaler (1982), shifted by 
   $E(B - V)$ = 0.04$\pm$0.01 
   ($E(U - B)$ = 0.029). The same $E(B - V)$ and $E(U - B)$ are then used 
   in the lower left and middle panel, respectively. Here, the isochrone is 
   superposed together with the ZAMS, both shifted by 
   ($V - M_V$) = 8.2$\pm$0.1. In 
   the right panel, the isochrone and ZAMS are shifted by the same 
   ($V - M_V$) and by $E(V - I)$ = 0.05. The fit is very good in all the 
   diagrams, and from this we derive a distance of 400$\pm$50 pc for
   NGC~1901.
   The reddening values we derived in different color indexes is
   compatible with a normal extinction law (Straizys 1992) toward the cluster.
   Starting from this point, we calculate the XYZ coordinates of NGC~1901
   assuming 8.5 kpc for the distance of the Sun to the Galactic
   Center. We find that X = 8.55$\pm$0.050, Y = 0.30$\pm$0.05 and Z
   =-0.220$\pm$0.05 kpc.

\section{Searching for fainter members}
    We use the results of the previous section to search for possible 
    fainter members for which no radial velocity information is
    available (see Table~4).

%
    \begin{figure}
     \centering
      \includegraphics[width=9cm]{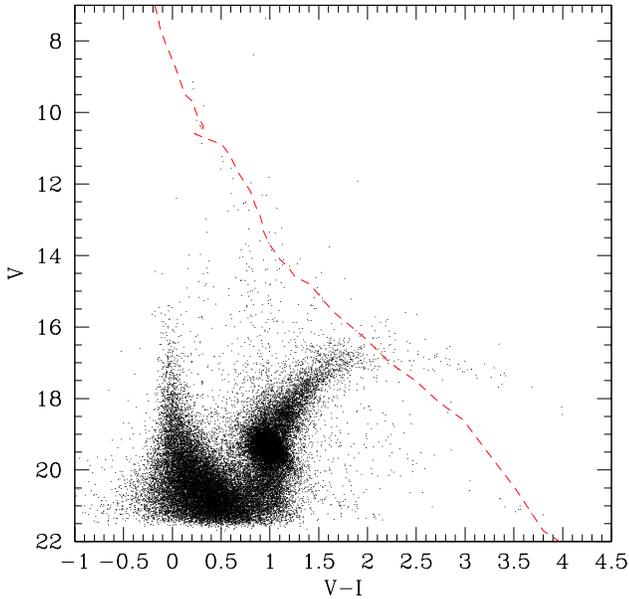}
      \caption{Color Magnitude Diagram: searching for NGC~1901 faint
	members. The dashed line is a ZAMS shifted by the distance
	modulus and reddening derived in Section~8.}%
    \end{figure}
%

    \noindent
    This is illustrated in Fig.~10, where the $V$ vs $V - I$ CMD is shown.
    Here, only the stars with errors in $V$ and $(V - I$) lower than 0.1 are 
    plotted. Basically, we search for faint members moving along the ZAMS, 
    which is displaced by the same reddening and apparent distance modulus 
    of above. We already have detected all the cluster members brighter than 
    $V$ = 14. Our objective is to look for stars close to the ZAMS
    (photometric probable members) which have as well 
    proper motion components compatible with the bulk of the cluster members. 
    This 
    strategy works down to $V$ = 16, which is the limit of the UCAC2 catalog. 
    In the magnitude range $14 \leq V \leq 16$ we found only 3 more cluster 
    possible members, namely $\#$ 34, 38, 93 (our numbering), but only
    one has compatible proper motion components. 
    Some other stars, although close to the ZAMS 
    ($\#$ 65, 70, 354 and 362), do  possess incompatible proper
    motions, too. Below 
    $V$ $\approx$ 16, the contamination of the RGB/AGB stars of the LMC makes 
    it impossible to search for NGC~1901 cluster members. It is however 
    possible that below the LMC RGB some M dwarfs members of NGC~1901 can be 
    present ($V$ $\approx$ 18), simply looking at their position in the CMD.
    However, the lack of any kinematic information prevent us from assessing 
    their membership in a reliable manner.

\begin{table*}
 \fontsize{8} {10pt}\selectfont
 \tabcolsep 0.25truecm
 \caption{NGC~1901 additional probable members. In the last column, PM
 indicates probable members, PNM probable non-members.}
 \begin{tabular}{cccccccccc}
 \hline
 \multicolumn{1}{c}{ID}         &
 \multicolumn{1}{c}{$\alpha$} &
 \multicolumn{1}{c}{$\delta$}        &
 \multicolumn{1}{c}{$U$}       &
 \multicolumn{1}{c}{$B$}       &
 \multicolumn{1}{c}{$V$}       &
 \multicolumn{1}{c}{$I$}       &
 \multicolumn{1}{c}{$\mu_{\alpha}$}       &
 \multicolumn{1}{c}{$\mu_{\delta}$}       &
 \multicolumn{1}{c}{Membership} \\
 \hline
 & hh:mm:sec & $o$:$\prime$:$\prime\prime$& mag & mag & mag& mag&
 mas/yr & mas/yr& \\
\hline
 34& 05:17:05.811&  -68:36:13.94&     14.38&   14.06&  13.26&  12.47&   19.8$\pm$6.4&      -3.7$\pm$6.1&PNM\\
 38& 05:17:51.366&  -68:26:08.28&     14.75&   14.40&  13.59&  12.69&    3.7$\pm$5.9&      21.3$\pm$5.9&PM\\
 65& 05:17:56.494&  -68:24:22.36&     15.54&   15.03&  14.12&  13.19&    7.4$\pm$5.9&      -8.8$\pm$5.9&PNM\\
 70& 05:20:02.775&  -68:24:43.23&     16.06&   15.36&  14.40&  13.36&    8.2$\pm$5.9&      23.4$\pm$5.9&PNM\\
 93& 05:17:07.433&  -68:34:40.54&     16.48&   15.77&  14.76&  13.64&   -2.0$\pm$6.4&      -9.2$\pm$6.3&PNM\\
354& 05:19:51.456&  -68:23:01.67&     18.57&   17.40&  16.14&  14.71&   -5.2$\pm$5.9&      29.2$\pm$5.9&PNM\\
362& 05:17:00.535&  -68:34:57.37&     18.68&   17.30&  16.05&  14.75&   10.8$\pm$5.9&       4.0$\pm$7.1&PNM\\
 \hline
 \end{tabular}
\end{table*}

\section{NGC~1901 Mass Function and dynamical status}
\begin{figure}
     \centering
      \includegraphics[width=9cm]{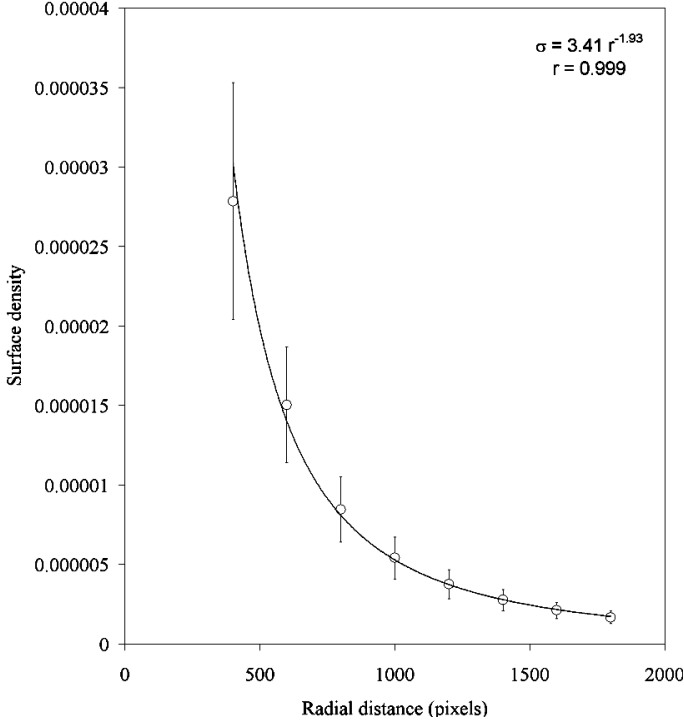}
      \caption{Star counts of assumed NGC~1901 members as a function 
               of radius.}%
      \label{profile}
    \end{figure}

    Table 2 suggests that NGC 1901 has currently 13 members with $V < 13$.
    If we study the radial distribution of these likely members we obtain
    Fig. \ref{profile}. The star density for a given value of the radius 
    in pixels is calculated counting all the sources (members) enclosed by 
    a circle of radius $r$ and dividing by the area ($\pi \ r^2$) to 
    obtain the surface density in stars/pixel$^2$. The final step in the 
    reductions was to calculate the statistical mean error of the star 
    density. If $n$ stars are expected in an area $A$, samplings of 
    independent areas of this size will give a Poisson distribution with 
    mean error $\sqrt{n}$ (King et al. 1968).  
    
    Figure \ref{profile} suggests that the effective radius of the cluster 
    (including half the members) is about 650 pixels or 3.1 arcmin. For the
    adopted value of the distance, this translates into an effective
    radius of 0.4 pc. The tidal radius is defined as the distance from the 
    cluster centre for which the relative acceleration becomes zero and it 
    corresponds to the maximum distance that a star can reach without 
    escaping from the cluster (e.g. Binney \& Tremaine 1987). At the tidal 
    radius the density distribution falls to zero. Following this criterion 
    and from Fig. \ref{profile}, the tidal radius can be estimated to be 
    $r_t$ = 1.0 pc (for a radius in pixels of 1800). The concentration 
    parameter:
    \begin{equation}
       C = \log \bigl(\frac{r_t}{r_c}\bigr) \,,
    \end{equation}
    is a value commonly used to measure how centrally concentrated a cluster 
    is. The core radius, $r_c$, is a well defined parameter of a King model 
    (King 1962, 1966) but from a theoretical point of view it is not clear
    how it should be defined in situations, like the present one, in which a 
    King model cannot be fit. On the other hand, the core radius as defined 
    observationally is the radius at which the density distribution has fallen 
    to just half the central density (for details see, e.g., Casertano \& Hut 
    1985). In our case, it gives a value of 416 pixels or 0.23 pc. Therefore, 
    the concentration parameter is rather low, about 0.64, that suggests a 
    relaxed system with a rather small halo.

    The current binary fraction of  NGC~1901 is rather high, 8 
    binaries out of 13 confirmed members. This translates into a binary
    fraction close to $62\%$. In the classical paper by Duquennoy and Mayor 
    (1991) about multiplicity in the Solar Neighborhood, they conclude 
    that two thirds of the studied systems were multiples. This figure is very
    close to the value found in the cluster studied here. OCRs are expected
    to contain a higher than average fraction of multiple systems. Our
    results for NGC~1901 are not fully conclusive in this respect as are
    also compatible with the characteristic value of the multiple fraction
    for the local field population. If our estimate for the present-day
    binary fraction of this object is correct and we assume that it is 
    a {\it bona-fide} OCR then simulations suggest that the primordial binary
    fraction for NGC~1901 was somewhat low, likely under 20$\%$. On the other
    hand, our result can also be interpreted as evidence of actual 
    incompleteness. It may be possible that a certain number of binaries
    remain hidden between the stellar contamination induced by the LMC
    and beyond $V = 13$. If true, this may contribute to underestimate the
    actual binary fraction. Fig. 10 appears to show a group of objects at
    $V \sim 21$ that may be cluster members.

    The present-day mass function of NGC~1901 appears to be quite unusual.
    If we assume that Fig. 10 can be considered as representative of the
    CMD for this object, it is clear that the number of K and M stars
    is rather low. Although it could be the result of incompleteness, our
    analysis suggests that the scarcity of low-mass cluster members could
    be real, the result of dynamical evolution or preferential evaporation
    of low-mass stars.

    The initial mass function (hereafter IMF) or frequency distribution of 
    stellar masses at birth (for a recent review on this subject see, e.g.,
    Chabrier 2003, also Scalo 1986) is a fundamental parameter for 
    studying the mass spectrum of open clusters. Salpeter (1955) used the 
    observed luminosity function for the Solar Neighbourhood and theoretical 
    evolution times to derive an IMF which may be approximated by a power-law: 
    \begin{equation}
       n(M) \propto M^{-\alpha} \,,  
    \end{equation}
    where $n(M)$ is the number of stars per unit mass interval. The value 
    of $\alpha$ is 2.35 for masses between 0.4 and 10.0 $M_{\odot}$. For
    the particular case of open star clusters, Taff (1974) found different 
    slopes in the range 2.50--2.74 depending on the concentration of stars 
    at the center of the cluster, the contrast of the cluster with the 
    surrounding stellar field --Shapley's (1933) and Trumpler's (1930) 
    classifications-- and the abundance of stars and range of brightness.
    The IMF in Taff's work can be described by $\alpha = 2.50$ for clusters 
    with $N \leq 100$ and $\alpha = 2.65$ for clusters with $N > 100$. 
    Note however that the canonical IMF parametrization constitutes a 
    two-part power-law IMF in the stellar regime (e.g. Pflamm-Altenburg \&
    Kroupa 2006):
    \begin{equation}
      n(M) = k \ \left\{ \begin{array}{lcl}
                        (M/M_{0})^{-\alpha_{1}} & & M_{0} \leq M < M_{1}\,, \\
                        (M_{1}/M_{0})^{-\alpha_{1}} (M/M_{1})^{-\alpha_{2}} & &
                         M_{1} \leq M < M_{max}\,,
                        \end{array} \right. \label{dis}
    \end{equation}
    where $M_{0}, M_{1}, M_{max}$ are equal to
    0.08, 0.5, 150.0 $M_{\odot}$, respectively, and $\alpha_{1}$, 
    $\alpha_{2}$ are 1.3, 2.3, respectively.

%
%
%

     \begin{figure}
      \centering
       \includegraphics[width=9cm]{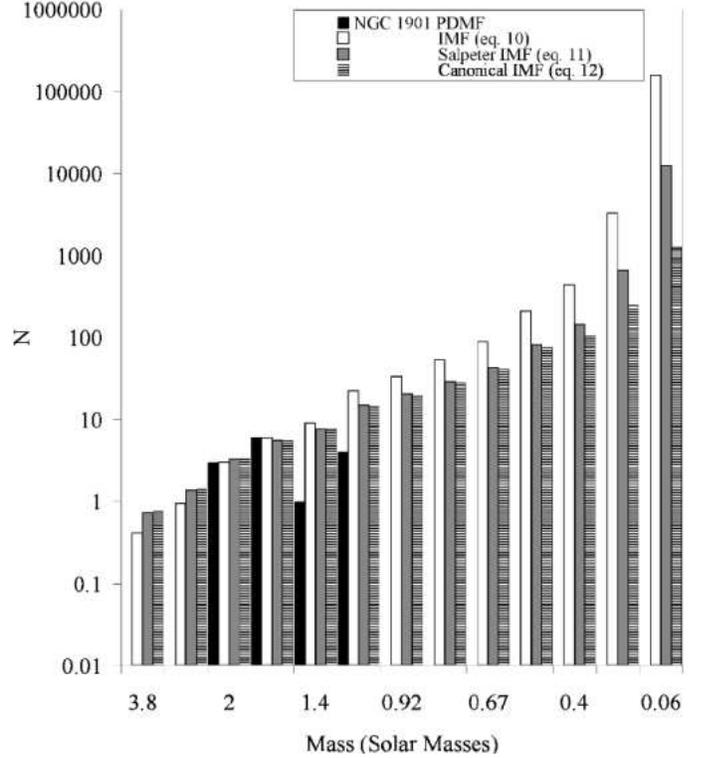}
       \caption{Observed present day mass function for NGC 1901 main
                sequence stars and various IMF fits. See the text for
                details.}%
       \label{imf}
     \end{figure}
%

    The present day mass function (hereafter PDMF) for main sequence stars 
    in NGC 1901 after Table~2 is displayed in Fig. \ref{imf}. It shows clear
    signs of incompleteness for spectral types later than A9. This could be
    a dynamical effect, the result of an increased escape rate for low-mass
    stars. It is well known that the dynamics of small and intermediate size
    clusters is dominated by relatively few heavy members. In the following
    we use the heaviest main sequence members of the cluster to estimate
    the IMF of NGC 1901. In these calculations, we assume implicitly that 
    the dynamical evolution of the cluster has not changed significantly the 
    population of A stars and that the actual number of these stars has
    remained relatively unchanged since the cluster was formed. If we use
    the mass intervals [1.6, 2.0] and [2.0, 3.0] $M_{\odot}$ (spectral types
    A9 to A5 and A4 to A0, respectively) to fit a simple power-law IMF we 
    obtain
    \begin{equation}
       n(M) =  (26 \pm 24) M^{-(3.1\pm1.1)} \,. 
    \end{equation}
    The result is displayed in Fig. \ref{imf}. This IMF produces an initial 
    cluster with a total mass of 10,600 $M_{\odot}$ and almost 30,000 members.
    This result clearly overestimates the initial number of low-mass stars
    in the cluster. On the other hand, if we fit the data to a Salpeter 
    IMF we obtain
    \begin{equation}
       n(M) =  (16.7 \pm 1.4) M^{-2.35} \,. 
    \end{equation}
    This IMF (see Fig. \ref{imf}) produces a cluster with an initial mass of 
    1,100 $M_{\odot}$ and about 2750 members. This is likely also an 
    overestimate. Our best result is obtained for a canonical two-part 
    power-law IMF:
    \begin{equation}
      n(M) = \left\{ \begin{array}{lcl}
                        (33\pm4) \ M^{-1.3} & & 0.08 \leq M < 0.5\,, \\
                        (16.3\pm1.4) \ M^{-2.3} & &
                         0.5 \leq m < 9\,,
                        \end{array} \right. \,. \label{best}
    \end{equation}
    The result is displayed in Fig. \ref{imf}. For this IMF, the initial 
    total mass of the cluster is 328 $M_{\odot}$ with about 820 initial
    members (estimated errors of 10\%). This is fully consistent with 
    results for the life-time of 750-1000 members simulated clusters 
    (e.g. de la Fuente Marcos, 1997).

    The present-day integrated visual magnitude of the cluster can be
    estimated using the expression
    \begin{equation}
       M_{V} - M_{V}^{\odot} =  -2.5 \ \log \frac{\Sigma L_{i}}{L_{\odot}} \,,
    \end{equation}
    where $M_{V}$ is the integrated absolute visual magnitude of the cluster,
    $M_{V}^{\odot}$ is the solar equivalent, $L = \Sigma L_{i}$ is the
    total luminosity of the cluster, and $L_{\odot}$ is the solar
    luminosity. The present day integrated absolute visual magnitude of 
    NGC 1901 including all the members in Table~2 is -0.4, far from the 
    value of the Hyades, -2.7, as expected. On the other hand, if we use our 
    best result for the IMF, the initial integrated magnitude could have been 
    as high as -1.4. 

\section{NGC~1901 motion}
  We derive here NGC~1901 kinematics by  considering the 6 {\it bona
  fide} single star members. We obtain

  \[  
  RV = 0.50\pm0.48 ~[km/sec]
  \]

  \noindent
  This is derived averaging two epochs RVs, when available, and then
  deriving the weighted mean of the five stars.

  \[
  \mu_{\alpha} \times cos \delta=  1.8\pm1.5 ~[mas/yr] 
  \]

  \[
  \mu_{\delta} = 10.7\pm1.7 ~[mas/yr]
  \]

   These values show a very good match with Baumgardt et al. (2000).

   In order to calculate the Galactic space velocity of this object and its 
   uncertainty, we average the heliocentric Galactic velocity components for 
   the five suspected single stars that have been considered as likely members 
   in Sect. 6, namely stars 2, 13, 14, 20 and 27. These calculations have been 
   carried out as described in Johnson and Soderblom (1987) for equinox 2000. 
   Our results are referred to a right-handed coordinate system so that the 
   velocity components are positive in the directions of the Galactic center, 
   U, Galactic rotation, V, and the North Galactic Pole, W. We use the value 
   of the parallax associated to the value of the distance determined in Sect. 
   8 and its error. Results are displayed in Table 5. They are based on 
   relative proper motions, not absolute ones.

\begin{table*}
 \fontsize{8} {10pt}\selectfont
 \tabcolsep 0.10truecm
 \caption{Space motions of suspected single NGC~1901 members.}
 \begin{tabular}{ccccccccc}
 \hline
 \multicolumn{1}{c}{ID}         &
 \multicolumn{1}{c}{Designation}        &
 \multicolumn{1}{c}{$\alpha$} &
 \multicolumn{1}{c}{$\delta$}        &
 \multicolumn{1}{c}{$\mu_{\alpha}$}       &
 \multicolumn{1}{c}{$\mu_{\delta}$}       &
 \multicolumn{1}{c}{U} &
 \multicolumn{1}{c}{V} &
 \multicolumn{1}{c}{W} \\
 \hline
& & hh:mm:sec & $o$:$\prime$:$\prime\prime$& mas/yr & mas/yr & km/sec
 & km/sec & km/sec \\
\hline
  2&       HD35294& 05:18:03.258&  -68:27:56.69&  1.0$\pm$1.2&       9.4$\pm$0.9&  -17.2$\pm$5.9 & -4.1$\pm$1.6  &  -4.2$\pm$2.2\\
 13&      HD269324& 05:18:29.529&  -68:27:13.81&  2.9$\pm$1.5&      11.3$\pm$2.0&  -21.3$\pm$9.3  & -2.6$\pm$1.7 &  -2.0$\pm$2.8\\
 14&      HD269334& 05:18:42.738&  -68:27:32.76&  1.9$\pm$2.9&       8.3$\pm$1.3&  -15.8$\pm$6.5  & -0.9$\pm$3.2 &  -1.0$\pm$4.7\\
 20&  GSC916200834& 05:19:05.214&  -68:30:45.75& -1.8$\pm$2.7&      11.7$\pm$1.5&  -21.8$\pm$8.3  & -0.1$\pm$3.0 &  -4.3$\pm$4.4\\
 27&  GSC916200626& 05:18:19.173&  -68:26:25.15&  0.7$\pm$2.4&       9.9$\pm$1.3&  -18.7$\pm$7.1  & -0.3$\pm$3.3 &  -2.0$\pm$4.1\\
 \hline
 \end{tabular}
\end{table*}

   From these values, we derive the Heliocentric reference frame velocity
   UVW, which turn out to be U = -19.0$\pm$1.2 km/s, V = -1.6$\pm$0.8 km/s 
   and W = -2.7$\pm$0.7 km/s. This yields a total velocity V$_T$ = 19.3 km/s.
   The standard deviations are $\sigma_U$ = 6.7 km/s, $\sigma_V$ = 2.9 km/s
   and $\sigma_W$ = 1.5 km/s. If only four stars are considered (star 27
   could be a long-period binary) our results are U = -19.0$\pm$1.5 km/s
   ($\sigma_U$ = 8.9 km/s), V = -1.9$\pm$0.9 km/s ($\sigma_V$ = 1.8 km/s), 
   and W = -2.9$\pm$0.8 km/s ($\sigma_U$ = 1.6 km/s), then V$_T$ = 19.3 km/s.
   These values are not in reasonable agreement with Eggen (1996), Dehnen
   (1998) and Chereul et al. (1999). In principle one may argue that using 
   relative proper motions may be a major factor in accounting for this lack 
   of agreement. SIMBAD lists proper motions from the Tycho Reference 
   Catalogue (Hog et al., 1998) for seven members (HD35294, HD35293, HD35462, 
   HD269338, HD269319, HD269310 and HD269301). The values provided by this
   catalogue agree very well with the ones used in this paper; therefore,
   the observed discrepancies have not been originated by the proper motions
   used. On the other hand, HD 35183 has parallax and proper motions in the 
   Hipparcos Catalogue (Perryman et al., 1997). Using the values provided by 
   this catalogue ($d$ = 820$\pm$584 pc, $\mu_{\alpha}$ = 1.07$\pm$1.03 mas/yr,
   $\mu_ {\delta}$ = 10.97$\pm$0.94 mas/yr) and our average value for the
   radial velocity, we obtain for this star U = -43$\pm$33 km/s, 
   V = -3$\pm$3 km/s and W = -6$\pm$7 km/s. This yields a total velocity 
   V$_T$ = 43.5 km/s. These numbers are much closer to the values usually 
   quoted in the literature for the Hyades stream (see Sect. 2) but they are
   the result of using a value of the distance that is twice the one found in 
   our study, even though they are still significantly different. We certainly
   believe that the value of the distance is the main source of discrepancies
   in this case. In any case, NGC~1901 is older than the Sirius supercluster
   (300 Myr) but younger than the Hyades supercluster (600 Myr).

\section{Conclusions}
We have presented a detailed photometric and spectroscopic study
of the stellar group NGC~1901 with the aim to provide an
observational template of a star cluster remnant to serve as example
for theoretical investigations of star cluster evolution
and dissolution.\\
The data we have acquired allow us to derive the following
conclusions:

\begin{description}
\item $\bullet$ we have identified 13 photometric, spectroscopic
and astrometric members;
\item $\bullet$ these stars identify a stellar group at 400 pc
from the Sun, and 400 million years old;
\item $\bullet$ within the errors of our spectroscopic campaign,
out of 13 stars, 8 turn out to be binaries,
which implies a binary fraction close to $62\%$; we suggest that
this is only a lower limit.
\item $\bullet$ we found a prominent lack of M dwarfs and interpret
this fact as the evidence that the group has lost most of its low
mass members;
\item $\bullet$  we cannot confirm the possible kinematic association
of NGC~1901 to the Stream I discussed in Eggen (1996); 
\item $\bullet$ in the
        light of numerical simulations (de la Fuente Marcos 1998),
this is compatible with NGC~1901
being what remains of a larger system initially made of
500-750 stars.
\end{description}

\noindent
Finally, we would like to emphasize that future studies should
be focused on better constraining the binary fraction of this
stellar group. Probably a few more epochs should be sufficient
to clarify the nature of all the doubtful cases.

\begin{acknowledgements}
    The work of GC has been supported by {\it Fundacion Andes}.
    In preparation of this paper, we made use of the NASA Astrophysics 
    Data System and the ASTRO-PH e-print server. This work made extensive 
    use of the SIMBAD database, operated at the CDS, Strasbourg, France. 
\end{acknowledgements}

\end{document}